\documentclass{article}

\usepackage[margin=1.20in]{geometry}
\usepackage[utf8]{inputenc} 
\usepackage[T1]{fontenc}    
\usepackage[english]{babel}
\usepackage{microtype}      
\usepackage{hyperref}       
\usepackage{url}            
\usepackage{lineno}
\usepackage{float}
\usepackage{authblk}
\usepackage{graphicx}
\usepackage[T1]{fontenc}
\usepackage{lmodern}
\usepackage{amsmath}
\usepackage{booktabs} 

\title{Cognitive networks identify the content of English and Italian popular posts about COVID-19 vaccines: Anticipation, logistics, conspiracy and loss of trust}

\author[1,*]{Massimo Stella}
\author[2]{Michael S. Vitevitch}
\author[1]{Federico Botta}
\affil[1]{Department of Computer Science, University of Exeter, UK}
\affil[2]{Department of Psychology, University of Kansas, USA}
\affil[*]{\small Corresponding author at: m.stella@exeter.ac.uk}

\begin{document}
\maketitle
\thispagestyle{empty}


\begin{abstract}

Monitoring social discourse about COVID-19 vaccines is key to understanding how large populations perceive vaccination campaigns. We focus on 4765 unique popular tweets in English or Italian about COVID-19 vaccines between 12/2020 and 03/2021. One popular English tweet was liked up to 495,000 times, stressing how popular tweets affected cognitively massive populations. We investigate both text and multimedia in tweets, building a knowledge graph of syntactic/semantic associations in messages including visual features and indicating how online users framed social discourse mostly around the logistics of vaccine distribution. The English semantic frame of "vaccine" was highly polarised between trust/anticipation (towards the vaccine as a scientific asset saving lives) and anger/sadness (mentioning critical issues with dose administering). Semantic associations with "vaccine," "hoax" and conspiratorial jargon indicated the persistence of conspiracy theories and vaccines in massively read English posts (absent in Italian messages). The image analysis found that popular tweets with images of people wearing face masks used language lacking the trust and joy found in tweets showing people with no masks, indicating a negative affect attributed to face covering in social discourse. A behavioural analysis revealed a tendency for users to share content eliciting joy, sadness and disgust and to like less sad messages, highlighting an interplay between emotions and content diffusion beyond sentiment. With the AstraZeneca vaccine being suspended in mid March 2021, "Astrazeneca" was associated with trustful language driven by experts, but popular Italian tweets framed "vaccine" by crucially replacing earlier levels of trust with deep sadness. Our results stress how cognitive networks and innovative multimedia processing open new ways for reconstructing online perceptions about vaccines and trust.

\end{abstract}

\section{Introduction}

Social media have given voice to millions of individuals. These massive digital audiences are increasingly being studied to better understand how socio-cognitive interactions create \cite{stella2018bots,mehler2020topic,bovet2018validation}, manipulate \cite{bessi2016social,ferrara2015quantifying} and promote \cite{gonzalez2021,varol2020journalists} specific perceptions about the online- and real worlds. On a more descriptive level, social media represent de-localised information systems where individual users share their own experiences and perceptions through language \cite{stella2021cognitive}. In this way, identifying cognitive features \cite{vitevitch2019can} of the language used on social media provides insight into how large audiences perceived and coped with specific events \cite{hills2019dark,stella2021cognitive}.

The present work used quantitative analyses to examine the language used on social media, in particular its emotional spectrum \cite{mohammad2010emotions} and semantic content \cite{fillmore2006frame}, to reconstruct popular perceptions about one specific event: the announcement of mass production of COVID-19 vaccines at the end of 2020. Taking inspiration from other recent works using social media as data for monitoring current perceptions and emotional responses to the coronavirus pandemic \cite{dyer2020public,yang2020analysis,stella2020lockdown,pierri2021vaccinitaly}, we here adopt the recent framework of cognitive network science \cite{siew2019cognitive,stella2019forma} in order to reconstruct the conceptual and emotional associations addressing COVID-19 vaccines. We focus our attention on popular messages on Twitter, with popularity being identified by the platform itself and corresponding to relatively high statistics of content sharing and liking. Our reconstruction of the key semantic frames reported in popular social media tweets represents a way to assess the online perceptions that reach massive audiences liked by up to 495,000 users (the population of a medium-sized city in the UK). Extracting and understanding these stances towards the COVID-19 vaccine is crucial in terms of identifying potential signals of distress \cite{fiorillo2020effects,aiello2020epidemic}, e.g. social users highlighting key challenges or denouncing issues with vaccine distribution that could promptly be solved once fully identified, or negative attitudes of closure \cite{jagiello2018bad,kalimeri2019human,mazzuca2021conceptual,hills2019dark}, e.g. conspiracy jargon that might hamper vaccination campaigns and have severe repercussions for pandemic containment. 

Building on the above necessity for quantitative investigations of socio-cognitive attitudes towards the COVID-19 vaccine, we provide more details about the cognitive aspects that characterise our approach and compare it against other data-driven works in the relevant literature \cite{kalimeri2019human,ferrara2015quantifying,pierri2021vaccinitaly,stella2020lockdown,montefinese2021covid}.

\subsection{Related work: Reconstructing cognitions and perceptions with complex networks}

The identification of people's views on something is known as stance detection in computer science and psycholinguistics \cite{kuccuk2020stance}. This task is key for understanding how conversations portray specific topics, such as individuals being in favour or against a given set of prescriptions about the COVID-19 pandemic. Studying stances through language has been historically performed through human intervention, which involves a person reading text, reconstructing syntactic and semantic associations between words in the text and then classifying the result. Human intervention is clearly not sustainable when dealing with thousands of interconnected stances as expressed in thousands and thousands of online social posts \cite{pierri2021vaccinitaly}. The advent of social media content gave voice to millions of internet users, a voice that could potentially report stances of relevance for understanding how real-world events are debated online \cite{stella2021cognitive}. Towards this direction, computer and data science recently developed numerous approaches and frameworks to tackle stance detection, mainly powered by machine learning and artificial intelligence, cf. \cite{mohammad2010emotions,mohammad2016sentiment,kuccuk2020stance}. Although machine learning is usually highly accurate in detecting whether a stance is positive or negative \cite{kalimeri2019human} or reconstructing the polarity and intensity of the sentiment expressed in language \cite{ferrara2015quantifying,mohammad2016sentiment}, it provides little information about the underlying structure of the stance. That is, such approaches are not able to understand how conceptual elements are entwined in a given stance in order to determine its overall meaning and emotional outline. Achieving more interpretable models of language processing and stance detection remains an open challenge \cite{kuccuk2020stance,hills2019dark}.

The present work merges machine learning with cognitive networks \cite{siew2019cognitive}, which are network representations of how linguistic knowledge is connected and processed within the human mind in a cognitive system known as the mental lexicon \cite{vitevitch2019can}. There are multiple ways to build networks out of texts, like using the relevance of words across paragraphs \cite{dearr2019paragraph} or co-occurrence networks to identify writing styles and authorship across manuscripts \cite{amancio2015probing}. However, in reconstructing a stance it is necessary to identify not only conceptual associations but also emotional trends and sentiment patterns \cite{kuccuk2020stance}. This combination is necessary in order to identify how individuals semantically framed specific concepts of social discourse (i.e. which associates surround a specific word), and what emotions revolve around each concept/idea. The inextricable connection between knowledge and affect recently led to the framework of textual forma mentis networks (TFMNs), cf. \cite{stella2020text}, which can extract syntactic, semantic, and emotional associations between words in text to create a knowledge graph that reconstructs the structure of the knowledge that authors embed in their posts. Networks like the ones used in the present work have successfully highlighted how students and researchers perceived STEM subjects \cite{stella2019forma}, how trainees changed their own mindset after a period of formal training \cite{stella2019innovation}, and have also identified key concepts in short texts \cite{stella2020text}. 

\subsection{Motivation: Cognitive networks operationalise semantic frame theory}

In cognitive science, semantic frame theory \cite{fillmore2006frame} indicates that meaning is attributed to individual concepts in language by means of syntactic/semantic relationships and further specified by words that are associated with that concept. In other words, the connotation denoted by an author to a concept might be reconstructed by checking which words were associated to it. For example, one text may frame the "gender gap" as a challenge that can be tackled by celebrating women's success in science, whereas another text may describe the "gender gap" in more pessimistic tones (cf. \cite{stella2020text}). Syntactic dependencies and semantic links thus provide key information for reconstructing the stance surrounding a given idea in terms of a network neighbourhood of concepts associated to a given word/idea coming from a mental lexicon \cite{vitevitch2019can}. Textual forma mentis networks operationalise semantic frame theory by identifying how concepts/words are associated to each other in sentences. In this network structure the first associates of a concept form a network neighbourhood that identifies how a concept was framed by authors, and which emotions populate that frame \cite{stella2021cognitive}. 

\subsection{Manuscript aims and outline}

This manuscript uses cognitive network science to reconstruct how popular tweets, available to hundreds of thousands of users, semantically framed and emotionally portrayed different aspects of COVID-19 vaccines. The Methods section contains details about the Twitter dataset, and the language and image processing methods implemented here. The Results section combines frequency- and networks-based analyses of key words in popular tweets together with quantitative inquiries of specific semantic frames reconstructed as networks. The Results include also behavioural and picture analyses reporting the emotions of content that was highly- or less-liked or -shared. The Discussion section links the current findings with the relevant research on COVID-19 and cognitive modelling. Emphasis is given to other works reporting analogous or synergistic results in terms of social media analyses, COVID-19 and human behaviour. 

\section{Methods}

\subsection{Twitter dataset}

This work relied on a main collection of 1962 unique popular tweets in English and 2413 unique popular tweets in Italian, gathered by the main author through Complex Science Consulting’s Twitter-authorised account ($@ConsultComplex$). Tweets were collected through ServiceConnect[] in Mathematica 11.3. Only tweets including the word "vaccine" or the hashtag \#vaccine were considered. The flag "Popular” in ServiceConnect[] gave access to trending tweets as identified by the Twitter platform.

Tweets were gathered between December 10 2020 and January 17 2021, a time window covering the early announcements of vaccines becoming available for mass vaccination and the subsequent discussion about the vaccination campaign.  Notice that the geographic location of tweets was not available in Mathematica 11.3, making it impossible to distinguish tweets based on their country of origin (e.g. US vs UK).
For each Tweet, statistics like the number of retweets and the number of likes (at the time of the query) were registered. English popular tweets were liked on average $20,000 \pm 48,000$ times, indicating a distribution of liked content critically skewed towards large values, and including tweets being liked up to 495,120 times. These English popular tweets were retweeted on average $3,000 \pm 6,000$ times with a single tweet being shared up to 57,821 times. Italian tweets registered lower values of liked content ($1,000 \pm 2,000$ with a maximum of 12,359 for a single tweet) and sharing ($150 \pm 200$ with a maximum of 2043 shares of a single tweet).

Twitter IDs and additional info like web links to pictures were also gathered and processed. 

After the temporary suspension of the AstraZeneca vaccine in several EU countries, including Italy, we gathered an additional set of 228 popular tweets in English and 180 popular tweets in Italian focusing on the keyword \textit{astrazeneca}.

\subsection{Language processing and network construction}

Text was extracted from each tweet in the dataset with the aim of building a knowledge graph of syntactic, semantic, and emotional associations between words, i.e. a textual forma mentis network (TFMN) \cite{stella2020text}. Emojis in tweets were translated in words by using Emojipedia (https://emojipedia.org/people/, last accessed 1 July 2020), which characterises individual emojis in terms of plain words. Hashtags were translated by using a simple overlap between the content of the hashtag without the \# symbol and English/Italian words (e.g., \#pandemic became “pandemic”). The resulting lists of words were then stemmed in order to get rid of word declination (e.g. "loving" and "love" representing the same stem "love"). Word stemming was performed by using WordStem[] in Mathematica 11.3 for English and SnowballC as implemented in R 3.4.4 for Italian (called through the RLink function in Mathematica 11.3). Stemming is particularly important for Italian, where nouns can be declined differently according to their gender (e.g., “dottoressa” and “dottore” both indicate the concept of a doctor). Stemming is important also in relation to the cognitive interpretation of a forma mentis network and knowledge representation in the human mind \cite{vitevitch2019can,doczi2019overview}. In fact, overwhelming evidence from psycholinguistics shows that different declinations of the same word do not alter the core meanings and emotions attributed to their stem \cite{doczi2019overview}. For instance, "loving" and "loved" both activate the same conceptual construct related to love in language processing by individuals. Hence, these words should be represented with the same lexical unit in a cognitive network representing human knowledge as derived from text. 

Knowledge representation was achieved through the building of a textual forma mentis network, whose main idea is to use machine learning for unearthing the complex network of syntactic relationships between words in sentences \cite{stella2021cognitive}. This network is not explicitly observed in the text (i.e., we do not see links between words when reading this or other texts), but is mentally reconstructed to associate the nouns, verbs, objects and specifiers in a sentence in order to figure out the meaning of a certain message. Textual forma mentis networks (TFMNs) are knowledge graphs enriched with cognitive perceptions about how massive populations associate and perceive individual words. Connections between lexical units/concepts are multiplex and indicate: (i) syntactic dependencies (e.g., in “Love is for the weak” the meaning of “love” is linked to the meaning of “weak” by the specifier “is for”) or (ii) synonyms (e.g., “weak” and “frail” overlapping in meaning in certain linguistic contexts). Syntactic dependencies were extracted from each sentence in tweets by using TextStructure[] in Mathematica 11.3, which relies on the Stanford NLP universal parser. Synonyms were identified by using WordNet 3.0 and its Italian translation \cite{miller1998wordnet}. The resulting syntactic/semantic network was enriched with emotional features attributed to individual words/nodes. Valence, arousal, and the emotions elicited by a given concept were attributed to individual words according to external cognitive datasets (see next subsection). In this way, TFMNs combine cognitive information about how individuals associate concepts eliciting different sentiment, excitement and emotions in texts.

The main English TFMN included 2190 words and 19534 links whereas the Italian TFMN contained 1752 words and 24654 links. The networks built in the aftermath of AstraZeneca's vaccine temporary suspension included respectively 410 words and 5953 links for English tweets, and 233 words and 2390 links for Italian tweets. "Vaccine" had a network degree \cite{siew2019cognitive} of over 800 in main English and Italian networks and of over 200 in networks based on popular tweets from the aftermath of vaccine suspension. Meaning modifiers like negation words (e.g. "not" or "no") were included in the network in order to keep track of meaning negation in emotional profiling. Words linked to negations were changed to their antonyms as extracted from WordNet 3.0 \cite{miller1998wordnet} and added to the semantic frame when computing emotional profiles.

\subsection{Cognitive datasets}

This study examined two datasets to reconstruct the emotional profile of language in texts: valence and arousal as coming from the psycholinguistic task implemented by Warriner and colleagues \cite{warriner2013norms}
and the Emotion Lexicon by Mohammad and Turney \cite{mohammad2010emotions}. Both datasets summarise how large populations of individuals perceive individual words, either by rating of pleasantness (valence) or excitement (arousal) or by listing which emotions are elicited by such words (e.g. "disease" elicits the emotion of fear). Valence and arousal act as coordinates in a 2D space mapping several human emotions. This mapping between language and emotions is known as the circumplex model \cite{posner2005circumplex} and it has been successfully used in several psycholinguistic investigations \cite{stella2020lockdown}. The emotional states reconstructed through the Emotional Lexicon were: Joy, Sadness, Fear, Disgust, Anger, Surprise, Anticipation and Trust. Although the first 6 emotional states are self-explanatory, the last 2 identify emotional perceptions of either projecting one's experience into the future (anticipation) or accepting norms and following behavioural codes imposed by others because of personal relationships or logical reasoning (trust) \cite{ekman1994nature}. 

We used the valence and arousal data of English words in order to build 2D density histograms identifying emotional trends in a given portion of language. As a linguistic baseline for emotional neutrality we adopted the interquartile range as computed from 13,900 English words in the Warriner et al. dataset \cite{warriner2013norms}. Clusters of words falling outside of the neutrality range indicate the presence of an emotional trend in language \cite{stella2020lockdown}. 

We used the Emotional Lexicon in order to count how many words $n_E(w)$ elicited a given emotion $E$ in a given semantic frame/network neighbourhood surrounding a concept $w$. We then compared each count against the expectation of a random null model drawing words at random from the overall emotional dataset. In each randomisation we drew  uniformly at random as many words as those eliciting for \textit{any} emotion present in the network neighbourhood. After repeating 500 random samplings, we computed a z-score for each emotion $E$, namely:

\begin{equation}
\centering
    z_E = \frac{n_{E}(w)- \langle n_E^{r}(w) \rangle}{\sigma^{r}_E(w)},
\end{equation}

where $\langle n_E^r(w) \rangle$ is the average random count of words eliciting a given emotion as expected in the underlying dataset (which features more words eliciting for some specific emotions and less concepts inspiring other emotions). $\sigma^{r}_{E}(w)$ is the standard deviation of the random counts. Z-scores higher than 1.96 (significance level of 0.05) indicate an excess of words eliciting a given emotion and surrounding the concept $w$ in the structure of social discourse. We plot emotional profiles as emotional flowers, where z-scores are petals distributed along 8 emotional dimensions. Petals falling outside of a semi-transparent circle, namely the rejection region relative to $z<1.96$, indicate a concentration of emotional jargon more extreme than expectation from the word-to-emotion mapping in common language (as coded in the sampled dataset and preserved by uniform random sampling).

The valence-arousal dataset was translated from English into Italian through a consensus translation using Google Translate, DeepL and Microsoft Bing. For the Emotion Lexicon, the authors used the automatic translations provided by Mohammad and Turney in Italian \cite{mohammad2010emotions}.

\begin{figure}[]
\centering
\includegraphics[width=14.5cm]{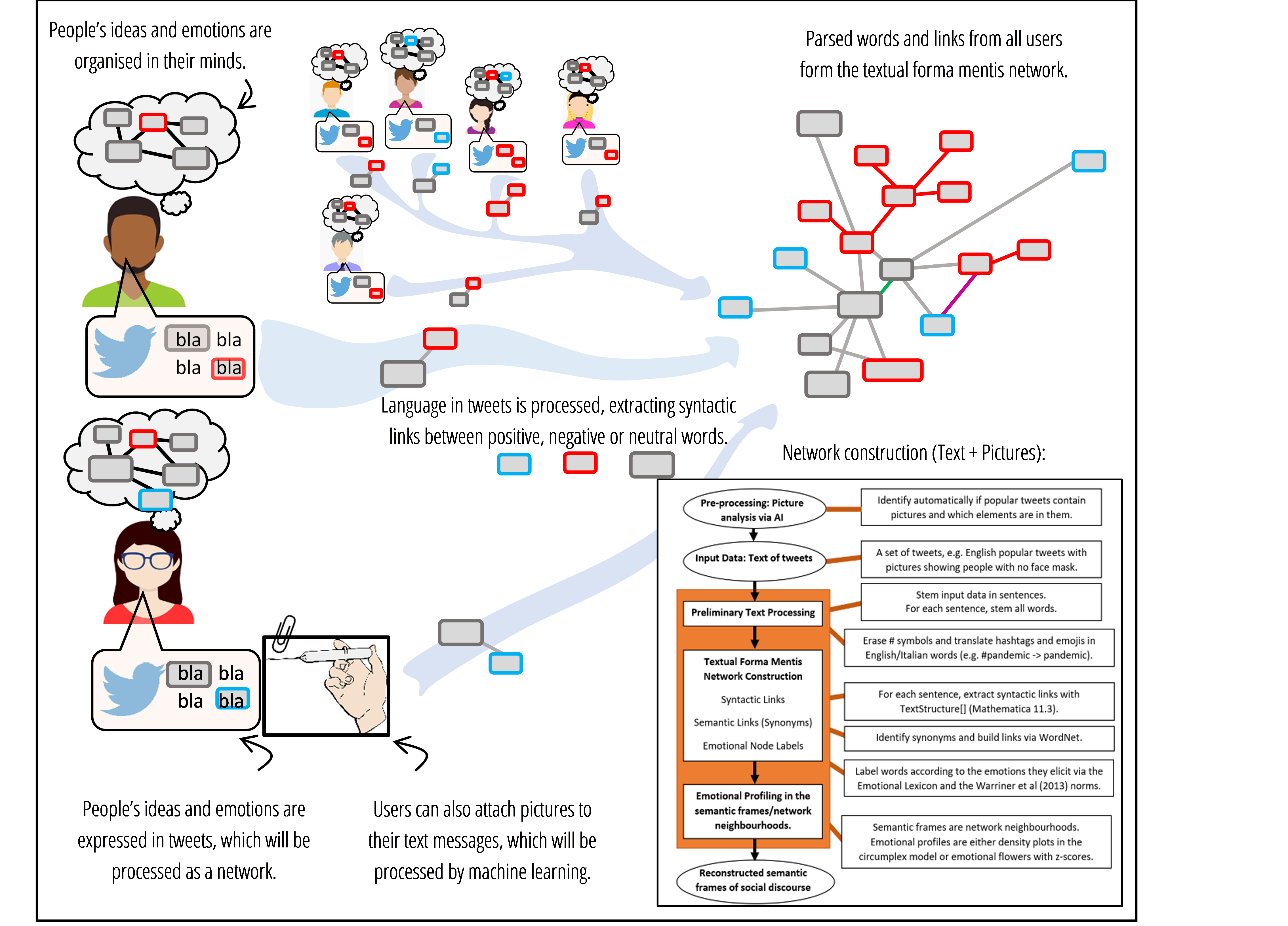}
\caption{Infographics about how textual forma mentis networks can give structure to the pictures and language posted by online users on social media. Semantic frames around specific ideas/concepts are reconstructed as network neighbourhoods. Word valence and emotional data make it possible to check how concepts were framed by users in posts mentioning (or not) pictures showing specific elements (e.g. people wearing a face mask). A flowchart with the different steps of network construction is outlined too.}
\label{fig:demonstrative}
\end{figure}

Figure \ref{fig:demonstrative} summarises the above steps for giving structure to the language and pictures posted in popular tweets. 

\subsection{Picture processing}
Tweets can often include images which are associated to the text. These can be used to support the emotional content shared in the tweet, and provide a visual medium which is complementary to text. Here, we analyse the image data in a variety of ways to complement the analysis of textual data, and to provide a further, deeper understanding of the emotional content shared on Twitter.

\subsubsection{Text extraction}
We download all images associated to English tweets and we process them using Google's Tesseract-OCR Engine via the Python-tesseract wrapper (\url{https://github.com/tesseract-ocr/tesseract}, Last Accessed: 22/03/2021). This uses a neural network based OCR engine to extract text from images. The resulting text from each processed image is then analysed using the language processing methods described above. It is important to highlight that not all images contain text, and in those cases the output of the OCR engine returns an empty string. Manual verification of a sample of the text extracted from the images has shown a good accuracy of the algorithm (above 95\%).

\subsubsection{Face and mask detection}
Face masks have been one of the trademarks of the COVID-19 pandemic, with the majority of countries worldwide introducing rules and recommendations on when and where face masks should be worn. Face masks have also often been a controversial topic, with polarised views from the general public on their perception in relation to personal freedom. As such, it is to be expected that a range of images associated to our tweets contain people wearing face coverings. This is of relevance to our analysis, since the public perception and polarisation about face masks will undoubtedly influence one's emotions about COVID-19 and vaccines.

However, the appearance and widespread use of face masks across the globe is mostly a recent phenomenon. Whilst face detection is a challenge which has been widely studied in the image processing community, detection of face masks has not been so prominent until recently. Detecting face coverings can be broken down into two different, sequential tasks: first, the algorithm has to detect the presence (and location) of faces within an image; then, for each detected face, the algorithm has to identify whether it is wearing a face mask. To analyse the images in our data set, we use a recently developed face mask detection algorithm made available via the facemask-detection Python package (\url{https://github.com/ternaus/facemask\_detection}, Last Accessed: 22/03/2021). This offers a pre-trained algorithm which carries out both the face detection step, and then assigns a probability to each detected face corresponding to the probability of there being a face covering. The face detection step uses a recently developed face detector, known as RetinaFace \cite{deng2019retinaface}. This algorithm uses state of the art deep learning techniques to output the location, in terms of a bounding box, of each face detected in an image. A key strength of RetinaFace is its ability to detect faces at various scales, where faces can be present both at the front as well as in the background of an image. On top of the RetinaFace layer, the face mask detection step uses a pre-trained set of deep neural networks to output a probability value for each face detected by the RetinaFace step. Probability values larger than 0.5 correspond to faces which the algorithm detected as wearing a face mask.

In our analysis, we process all images and, for each image, we collect the number of detected faces by the RetinaFace layer, as well as how many of those faces have been assigned a probability larger than 0.5 of having a face mask. Note that we include in this analysis all images, regardless of whether they contain any text or not. This is because we have observed that some images contain text overlayed on top of a normal image, which in some cases contains faces and masks. Therefore, the face mask detection step is applied to all images in the data set of English tweets.

Finally, it is worth highlighting that, whilst manual inspection of the results of the face and mask detection step show a very good accuracy, there are undoubtedly some cases in which this method fails to identify all faces or masks correctly. Whilst to be expected, such a limitation must be kept in mind when drawing conclusions from the results of our analysis.

\subsubsection{Dominant colour analysis}

Another key component of the visual aspect of images associated to tweets is that associated to colours. Colour analysis of images posted to Instagram has revealed a link between Hue, Saturation and Value (HSV) and individuals with depression \cite{reece2017instagram}. This suggests that images may, to an extent, reflect the emotional and well-being status of individuals who choose to share those images online. Here, we extract the dominant colour values in terms of the hue value. The hue value represents the colour on the light spectrum, with low values representing red and large values representing blue and purple. To extract a single hue value from each image, we perform a two-step analysis on each image which does not contain textual data. First, we run the k-means clustering algorithm on the HSV values of each pixel for each image. We then extract the centroid of the largest cluster found, and consider the corresponding HSV values as the dominant values of the image. Note that this is only an approximation of the dominant colour, and we use $k=5$ in each image. After this initial step, each image is represented by its dominant HSV values. Visual inspection of the resulting dominant HSV values across all images analysed indicates the presence of two strong clusters in the hue component, with one cluster centred on low values of hue (in the red spectrum) and a second cluster centred on large values of hue (in the blue spectrum). Based on this finding, we perform a second clustering step, using the k-means clustering algorithm with $k=2$ to group the images in two clusters: one with dominant hue values in the red area; and one with dominant hue values in the blue area. More detail on this are provided in the Supplementary Information figures.

\section*{Results}

This section outlines the results of the analysis of popular tweets in terms of: (i) prominent concepts in social discourse captured by frequency of occurrence and network centrality, (ii) focus on the emotional-semantic frames of "vaccine" and "vaccino" in social discourse, (iii) other semantic and emotional frames of prominent concepts in social discourse, (iv) behavioural comparisons of tweet sharing and liking depending on the emotional profile of posts, (v) picture-enriched analysis of online language.

\subsection{Prominent concepts in social discourse captured by frequency of occurrence and network centrality}

This part of the study focused on identifying the key ideas reported in social discourse about the COVID-19 vaccine. Table 1 provides the 20 top-ranked concepts identified through word frequency and degree in the TFMN. Word frequency identifies how many times a single word was repeated across popular tweets and were potentially read by users. Degree counts how many different syntactic/semantic associations were attributed to a given concept, and captures in the textual forma mentis network semantic richness, i.e. the number of connotations and semantic associates attributed to a single word \cite{stella2020lockdown,siew2019cognitive}.

As highlighted in Table 1 (left), English popular tweets featured mostly jargon relative to the idea of "people receiving their first dose of vaccine". This pattern was consistent between the ranks based on word frequency and semantic richness/network degree, respectively. These prominent words and additional key jargon related to the semantic sphere of time (like "week", "now" and "when") together indicate a social discourse dominated by a projection into the future, relative to the logistics of vaccine distribution. Network degree identified also Trump as a key actor of popular tweets. Differently from word frequency, semantic richness highlighted how "workers" and "work" were prominently featured in popular tweets.

The Italian social discourse also featured key words related to people receiving their first dose of the vaccine. Key actors of social discourse in the Italian twittersphere were Pfizer and Moderna. Italian users mentioned medical jargon (e.g. "doctor", "virus", "effects") more prominently than English speakers, in terms of both semantic richness and frequency. As in English, Italian discourse was also strongly dominated by words related to the semantic sphere of time, including prominent words like "time", "hour" and "day".

The above rankings indicate how social discourse in popular tweets about the COVID-19 vaccine were prominently projected towards future plans for dose distribution. According to the above simple semantic analysis, it can be postulated that the specific semantic frame of "vaccine" (and "vaccino" in Italian) also should be populated by emotions like anticipation into the future.

\begin{table}[ht!]
    \centering
    \begin{tabular}{ccc}
 \textbf{Rank} & \textbf{Degree} & \textbf{Frequency} \\
 \hline
 1 & \text{vaccine} & \text{vaccine} \\
 2 & \text{will} & \text{we} \\
 3 & \text{get} & \text{covid} \\
 4 & \text{people} & \text{will} \\
 5 & \text{dose} & \text{get} \\
 6 & \text{take} & \text{people} \\
 7 & \text{say} & \text{first} \\
 8 & \text{receive} & \text{now} \\
 9 & \text{first} & \text{all} \\
 10 & \text{new} & \text{dose} \\
 11 & \text{make} & \text{take} \\
 12 & \text{trump} & \text{coronavirus} \\
 13 & \text{govern} & \text{million} \\
 14 & \text{out} & \text{when} \\
 15 & \text{worker} & \text{new} \\
 16 & \text{distribute} & \text{need} \\
 17 & \text{work} & \text{health} \\
 18 & \text{million} & \text{after} \\
 19 & \text{week} & \text{rollout} \\
 20 & \text{need} & \text{virus} \\
\end{tabular}
\quad
\quad
\quad
\quad
\begin{tabular}{ccc}
 \textbf{Rank} & \textbf{Degree} & \textbf{Frequency} \\
 \hline
 1 & \text{vaccino} & \text{vaccino} \\
 2 & \text{prima (first)} & \text{dose} \\
 3 & \text{dose} & \text{contro(against)} \\
 4 & \text{stato (state)} & \text{pi{\` u} (plus)} \\
 5 & \text{tutti (all)} & \text{covid} \\
 6 & \text{contro (against)} & \text{prima (first)} \\
 7 & \text{fatto (fact)} & \text{pfizer} \\
 8 & \text{chi (who)}  & \text{italia} \\
 9 & \text{persona (person)} & \text{tutti (all)} \\
 10 & \text{casi (cases)} & \text{tempo (time)} \\
 11 & \text{dati (data)} & \text{moderna} \\
 12 & \text{virus} & \text{fatto (fact)} \\
 13 & \text{medico (doctor)} & \text{solo (only)} \\
 14 & \text{tempo (time)} & \text{virus} \\
 15 & \text{prendere (take)} & \text{ora (now/hour)} \\
 16 & \text{parte (part)} & \text{giorno (day)} \\
 17 & \text{morti (deaths)} & \text{ansa} \\
 18 & \text{paese (country)} & \text{oggi (today)} \\
 19 & \text{passare (transit)} & \text{dati (data)} \\
 20 & \text{arrivare (arrive)} & \text{effetti (effects)} \\
\end{tabular}
    \caption{Top-20 key concepts in the English (left) and Italian (right) corpora. Words are ranked according to their degree in textual forma mentis networks and their frequency in the original tweets. Italian words were translated in English for an easier visualisation.}
    \label{tab:my_label}
\end{table}

\subsection{Semantic frames and emotional profiles of "vaccine" in popular tweets}

\begin{figure}[]
\centering
\includegraphics[width=4cm]{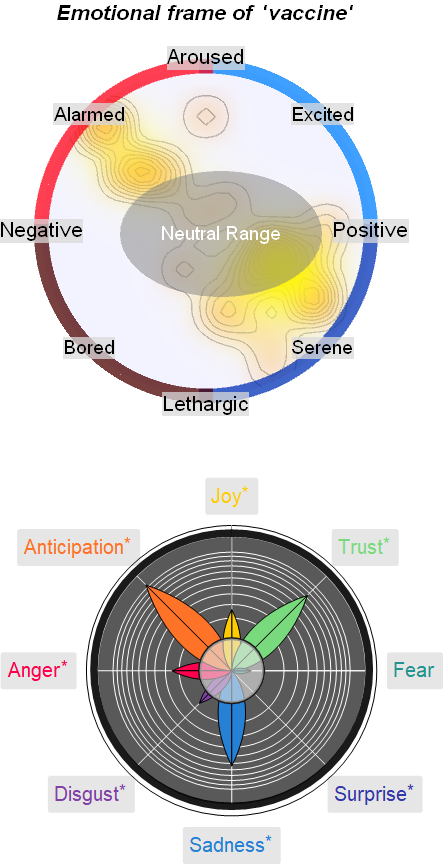}
\includegraphics[width=10cm]{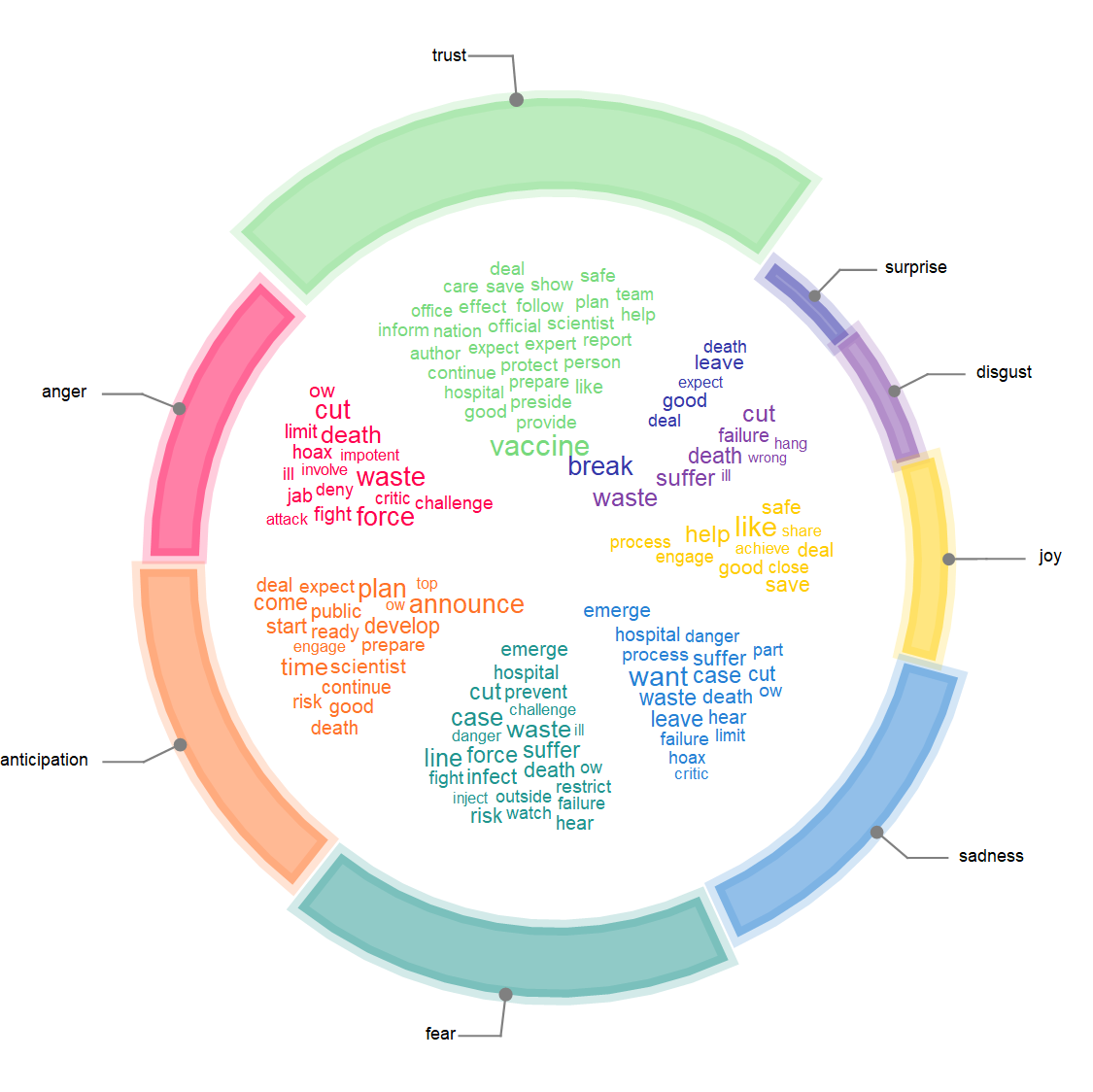}
\includegraphics[width=4cm]{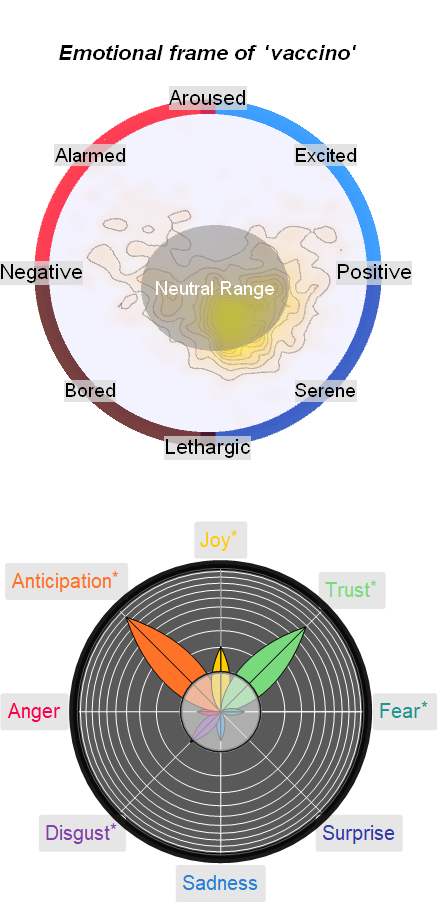}
\includegraphics[width=10cm]{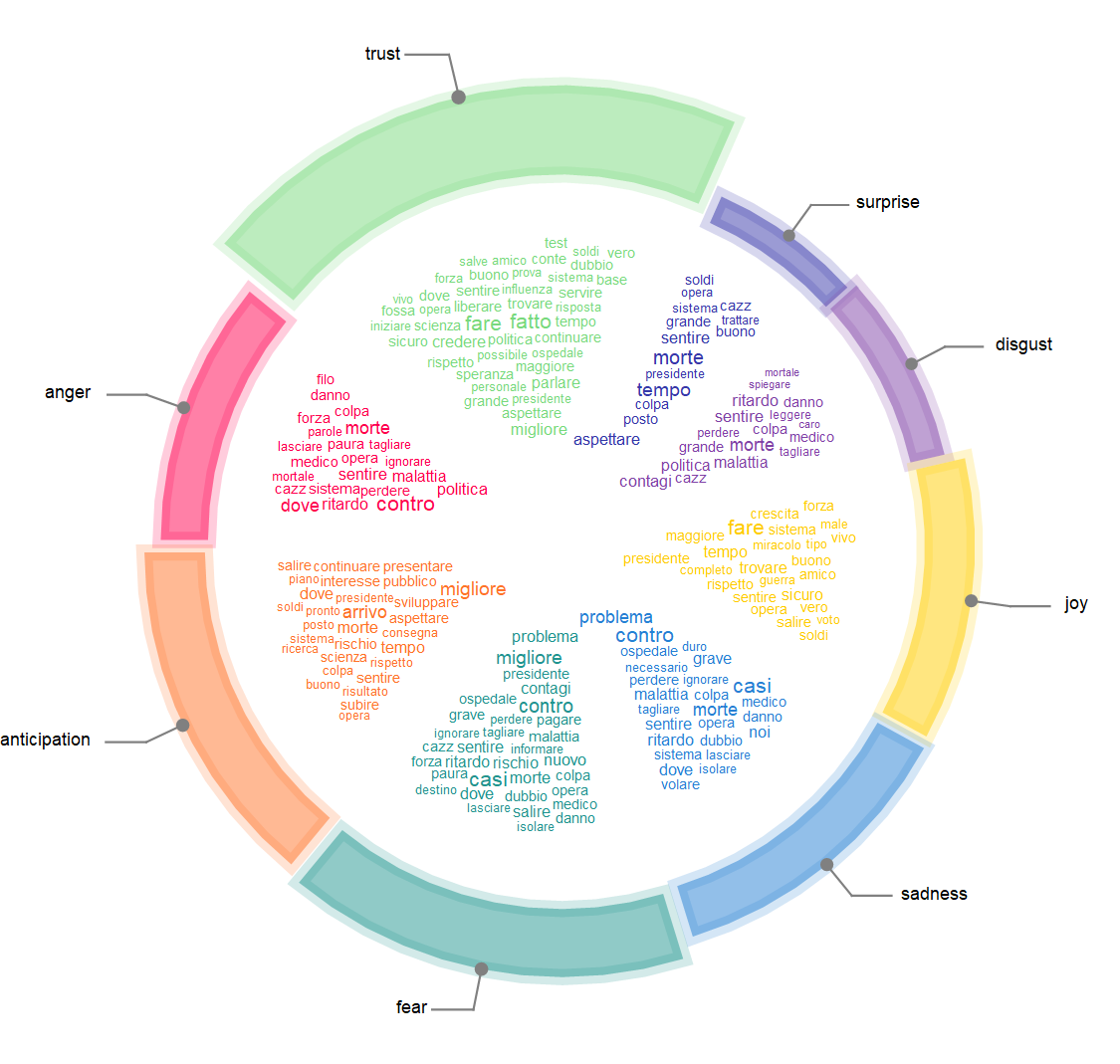}
\caption{Emotional analysis and word clouds of concepts in the semantic frame of "vaccine" (in English) and "vaccino" (in Italian). The circumplex model indicates how the neighbors of vaccine/vaccino populate a 2D arousal/valence space. The emotion flower indicates an excess of emotions detected in the semantic frame compared to random expectation. The sector chart reports the raw fraction of words eliciting a certain emotion. The word cloud reports the top 10\% concepts with the highest degree centrality and associated with vaccine. Words are distributed according to the emotions they elicit.}
\label{fig:vacci}
\end{figure}

Figure \ref{fig:vacci} focuses on the semantic frame surrounding "vaccine" in both English (top) and Italian (bottom) popular tweets. Three representations are compared: (i) a word cloud of words in the neighbourhood of "vaccine" with size proportional to their degree in the overall TFMN and a sector chart counting the proportion of words eliciting a certain emotion, (ii) an emotional flower indicating excess of emotions in the same neighbourhood as z-scores/petals (see Methods), and (iii) a circumplex model of emotions plotting a 2D valence/arousal histogram of words populating the neighbourhood of "vaccine". Both the circumplex model and the emotional flower (models relying on different datasets) agree in indicating a polarised emotional profile of "vaccine" in popular tweets, concentrating around both positive/calm and negative/alerted emotional states. As reported in the emotional flower, anticipation into the future was the strongest emotional state populating the semantic frame of "vaccine". However, the petals/z-scores falling outside the rejection region (white circle) indicate also a concentration of words eliciting trust and joy, but also anger, disgust, and sadness that is higher than random expectation.

As described in the Introduction, TFMNs enable a direct access to the semantic frame surrounding key ideas in social discourse. The above results indicate that popular tweets were indeed mostly dominated by words related to future events (anticipation), as indicated also by the above prominence analysis. However, the framing of vaccines in tweets was not emotionally uniform, but rather strongly polarised around alarming and more positive/calm tones. The word cloud in Figure \ref{fig:vacci} (top) identifies how emotions where associated with different concepts.

\subsection{Semantic frames and emotional profiles of other prominent concepts of social discourse}

The rankings based on degree/semantic richness and frequency of words in social discourse show that a key theme of popular tweets was "people getting a dose of vaccine". This key topic and the overall dominance of anticipation into the future, as reported in the previous section, both underline how relevant the logistics of vaccine distribution in the health system was in the considered COVID-19 tweets. This makes it important to investigate how concepts like "distribute" and "health" were framed by social discourse.

\begin{figure}[]
\centering
\includegraphics[width=14.5cm]{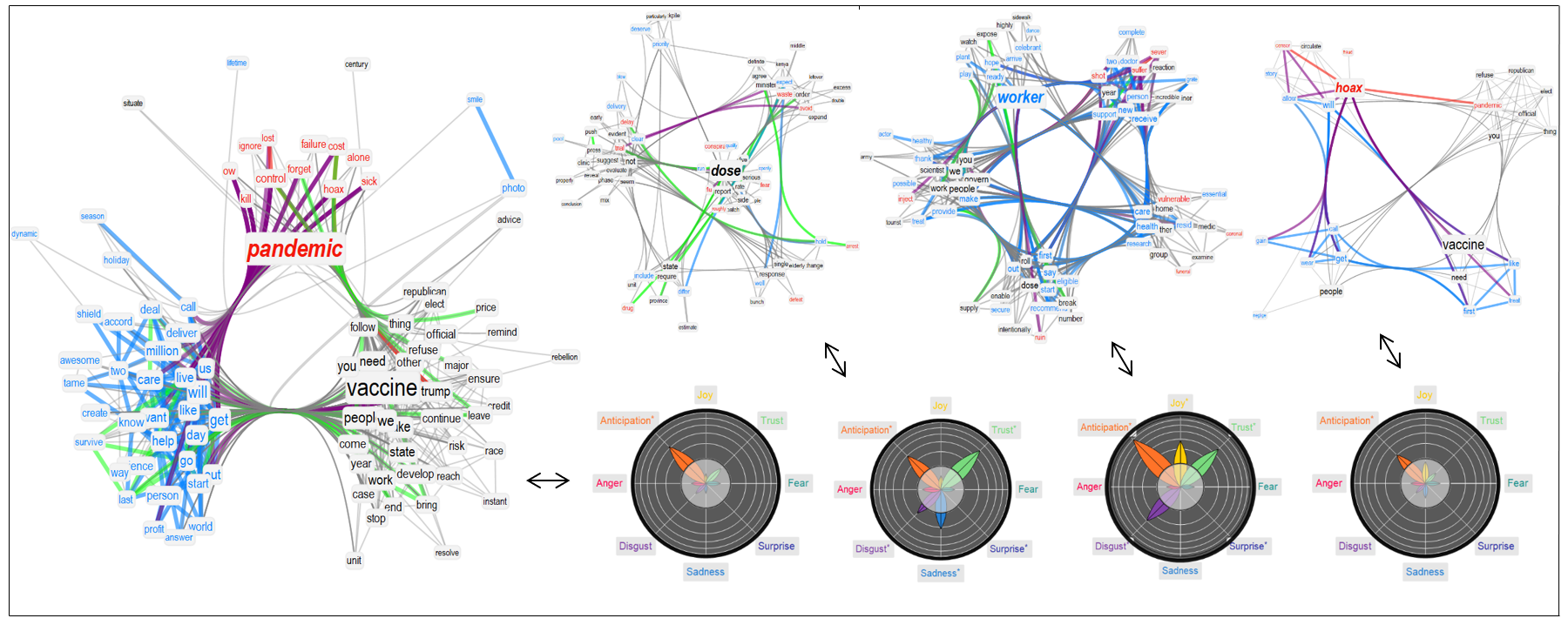}
\includegraphics[width=14.5cm]{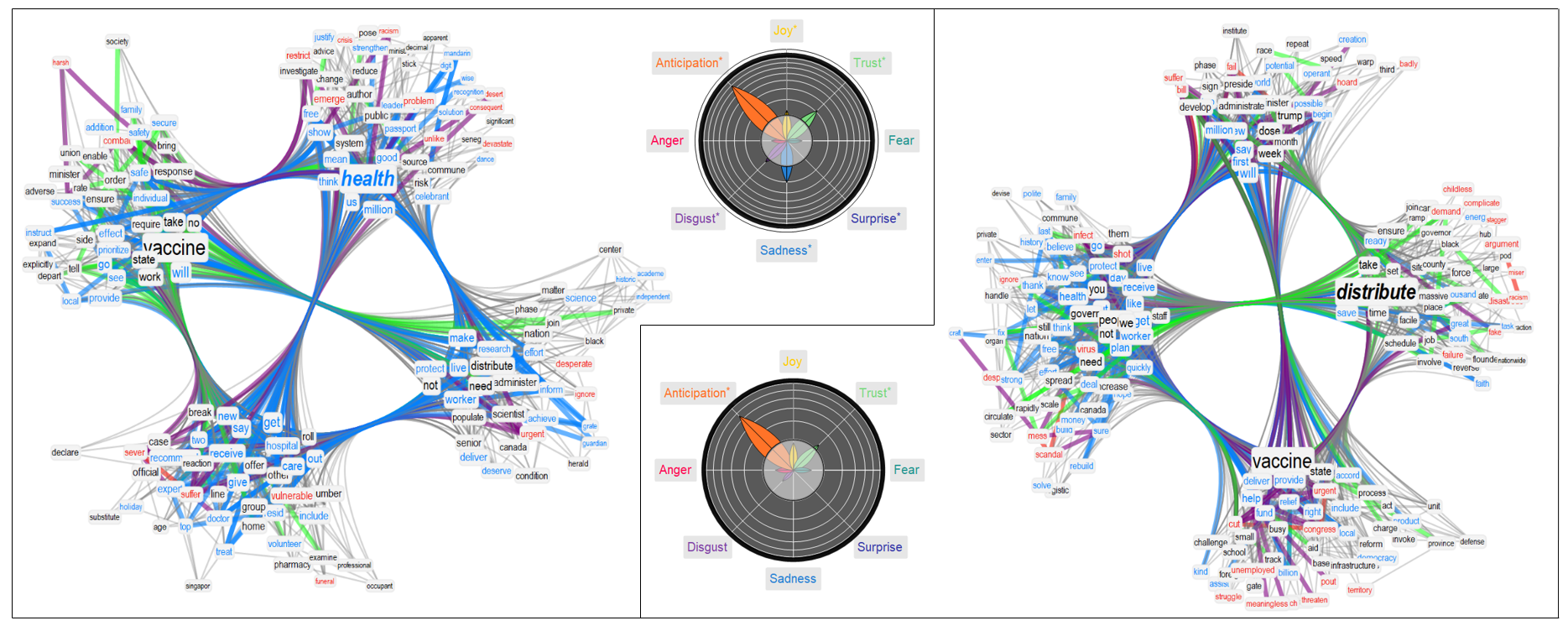}
\caption{TFMNs capturing conceptual associations in social discourse around "pandemic", "dose", "worker" and "hoax" (top) and around "health" and "distribute" (bottom). Positive (negative) concepts are cyan (red). Neutral concepts are in blue. Associations between positive (negative) concepts are highlighted in cyan. Purple links connect concepts of opposite valence. Green links indicate overlap in meaning. The emotional flowers indicate how rich the reported neighbourhoods are in terms of emotional jargon. Petals falling outside of the inner circle indicate a richness that differs from random expectation at $\alpha=0.05$. Each ring outside of the circle corresponds to one unit of z-score.}
\label{fig:pandemic}
\end{figure}

Figure \ref{fig:pandemic} (bottom) reports the semantic frame surrounding "health" (bottom left) and "distribute" (bottom right). In both the visualisations, semantic frames are identified as network neighbourhoods and organised in communities as identified by the Louvain algorithm \cite{blondel2008fast}. Network communities correspond to more tightly interconnected clusters of words, reflecting different semantic aspects of a given frame \cite{siew2019cognitive}. The network visualisations are complemented by emotional flowers. Discourse surrounding "health" was mostly dominated by syntactic/semantic associations with other positive jargon, featuring both: (i) actors of the health system (e.g. doctors, hospitals, volunteers), and (ii) aspects of vaccine delivery and administration. Popular tweets importantly linked vaccine distribution with vulnerable groups and mentioned the urgency of suitable plans for administering the vaccine despite the current crisis. 

Importantly, popular tweets also drew a conceptual association between "health" and "racism", making a point about the necessity of fair measures of health provision. The overall emotional profile of all the above aspects was dominated by anticipation but included also sadness (related to the difficulties of the current crisis) and trust towards the health system, its actors and its supporters, like countries, nations and science.

Popular tweets were less positive when framing the specific concept of "distribute", whose semantic frame was mostly populated by anticipation into the future. The forma mentis network highlighted many negative associations indicating a potentially challenging and disastrous management of the limited available resources  (e.g. "failure", "badly", "hoard", "suffer", "demand", "scandal"). Positive associates were generally confined to the semantic area of product delivery and democratic, quick management of resources. Institutions, administrations and nations (e.g. "president", "trump", "administrate") were tightly connected with concepts related to the semantic spheres of speed and time (e.g. "week", "month", "speed", "warp"). This represents additional evidence that popular tweets about the COVID-19 vaccine underlined the need for a quick administering of vaccine doses. 

A shift into the future is present also in other semantic frames, see Figure \ref{fig:pandemic} (top). Whereas previous investigations reported semantic frames for "pandemic" filled with negative emotions \cite{stella2020lockdown}, in the popular tweets analysed here the current pandemic was contrasted with overwhelmingly positive jargon, featuring concepts like "care", "create", "live" and "shield". These conceptual associations balanced out negative emotions and provided an overall frame of associates for "pandemic" mostly devoid of any emotion except for anticipation into the future.

Negative emotions like disgust or sadness were found in the associates of "dose" and "worker", together with more positive emotional states like trust and joy. More in detail:
\begin{itemize}
    \item The associations attributed to "dose" identified aspects like "delay", "trial", "waste", "fear" and "conspiracy", highlighting expressions of concerns about the validity of a dose of vaccine;
    \item Sadness around "workers" had as semantic associations "vulnerable", "expose", "funeral", "essential", "suffer" and "severe", indicating how popular tweets highlighted the importance for exposed workers to receive a vaccine. 
    \item The above trend co-existed with positive emotions originating from celebratory jargon ("thanks", "celebrate") identifying the importance of workers during the pandemic.
\end{itemize}

Whereas Italian popular tweets did not feature jargon related to conspiracy theories, English popular tweets provided a rather highly clustered network neighbourhood for "hoax", devoid of negations of meaning and featuring mostly jargon related to the future. Associations of "hoax" with ideas like "censor", "pandemic" and "vaccine" indicate a concerning portrayal of conspiracy theories within the considered sample of popular tweets. This represents quantitative evidence that conspiracy theories revolving around the COVID-19 vaccine were capable of reaching large audiences online through highly shared and liked (i.e. popular) tweets.

\subsection{User behavioural trends on Twitter based on emotions}

\begin{figure}[]
\centering
\includegraphics[width=13cm]{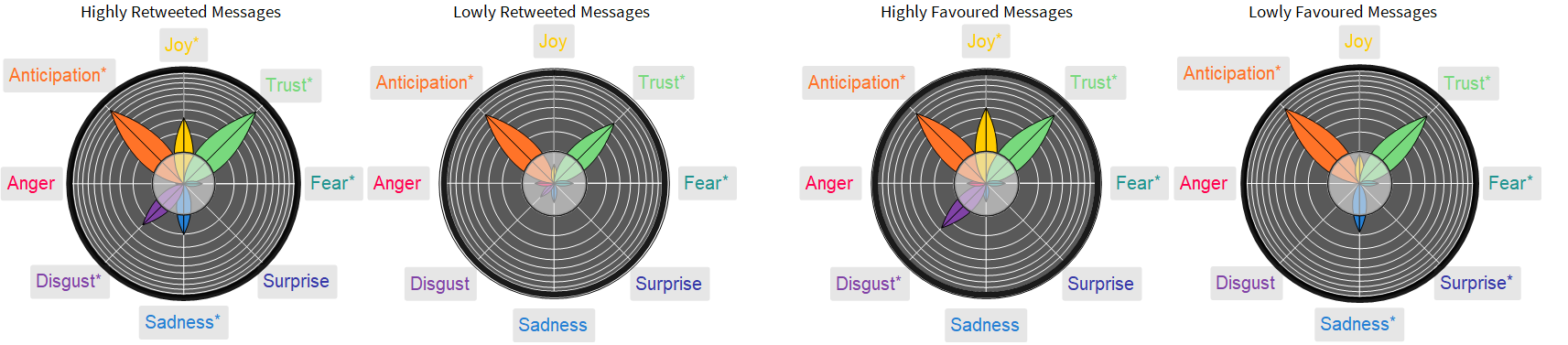}
\includegraphics[width=13cm]{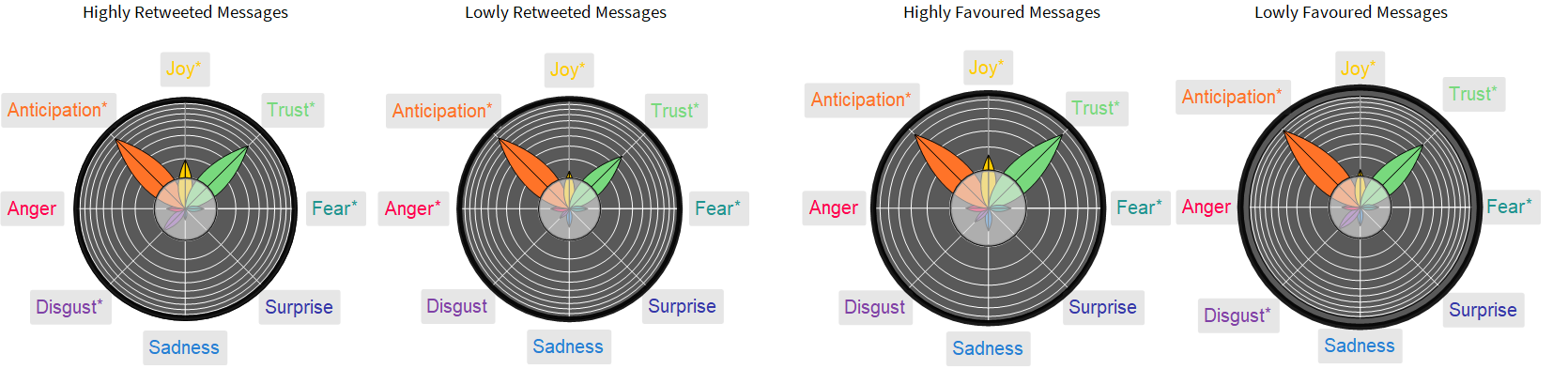}
\caption{Multi-language analysis of the emotional profiles of highly-/less retweeted (left) or liked (right) tweets in English (top) and in Italian (low). Petals indicate z-scores and are higher than the threshold of 1.96 when falling outside of the semi-transparent circle.}
\label{fig:behaviour}
\end{figure}

To test how users reacted to emotional content in popular tweets we studied the emotional profiles of highly-/less-retweeted and liked messages. Notice that these distinctions were based on retweet or like counts being higher or lower than their respective medians. Figure \ref{fig:behaviour} reports the emotional profiles of highly-/less retweeted (left) or liked (right) tweets in English (top) and in Italian (low). In English, highly retweeted or liked tweets contained language with an emotional content drastically different to the one embedded in less shared or liked content. An excess of sadness, joy, and disgust characterised highly retweeted text messages, emotions absent in popular yet less frequently retweeted content. The measured levels of trust and anticipation remained high across all the considered classes.

These results indicate that the behavioural strategies behind content sharing and the tendency for users to retweet content displayed an emotional interplay. Trust and anticipation, the strongest emotions surrounding "vaccine", did not change significantly between highly/less re-tweeted and liked content. Instead, positive emotions of high arousal (i.e. joy) or negative emotions eliciting risk-averse behaviour (i.e. disgust or sadness) corresponded to a boost of content spreading in English. Interestingly, in another language, i.e. Italian, only a small difference in joy was detected. Notice how, with like counts, sadness characterised less liked popular tweets and was not found in highly liked messages. This indicates a tendency for users not to like sad content while still actively engaging with it through retweeting. 

\subsection{Processing together pictures and text: Colours, people and face masks}

In addition to text processing, we enriched our analysis with an investigation of the multimedia content promoted in popular tweets. 

An analysis reading the text of pictures included in popular tweets (see Methods) identified portions of language sharing the very same emotional content of the shorter text of tweets themselves. This finding indicates an overall consistency of language between the header of a tweet and the content of the picture attached to it. 

With no difference detected in the text reported in pictures, we focused our attention on pictures portraying negligible, e.g. a single word, or no text at all. An analysis of the hue values identified mainly two predominant colours in these pictures, as contained in popular tweets, namely hue values in the red region of the spectrum and hue values in the blue region (see Supplementary Figure 1). Whereas the circumplex model identified a polarisation of emotions being present across all tweets and ranging between calmness to alarm, emotional profiling through the Emotional Lexicon identified a lower level of anticipation into the future in the language describing specifically pictures with a predominant blue colour (see Supplementary Figure 2). A human coding of pictures revealed that blue was the main colour for the backgrounds and foregrounds in pictures displaying people being vaccinated. Hence the observed decrease in anticipation indicates that when describing the specific event of vaccination, language becomes less projected into the future. 

In order to better investigate how language and pictures are entwined, we performed a specific content analysis based on machine vision and focusing on the portrayal of people and pandemic-related objects. Since the detection of syringes or vaccine vials would be problematic, we focus on objects more tightly connected to people like face masks.

\subsection{Investigating the language of tweets with pictures of people wearing, or not, face masks}

We built three textual forma mentis networks each based on one of the following corpora of tweets: (i) posts including pictures of no people, (ii) posts including pictures of people wearing no face masks and (iii) posts including features of people wearing face masks. Note that we include in this third category all tweets for which an image contains at least one face mask, even though there may be other faces without a mask.

\begin{figure}[]
\centering
\includegraphics[width=3.cm]{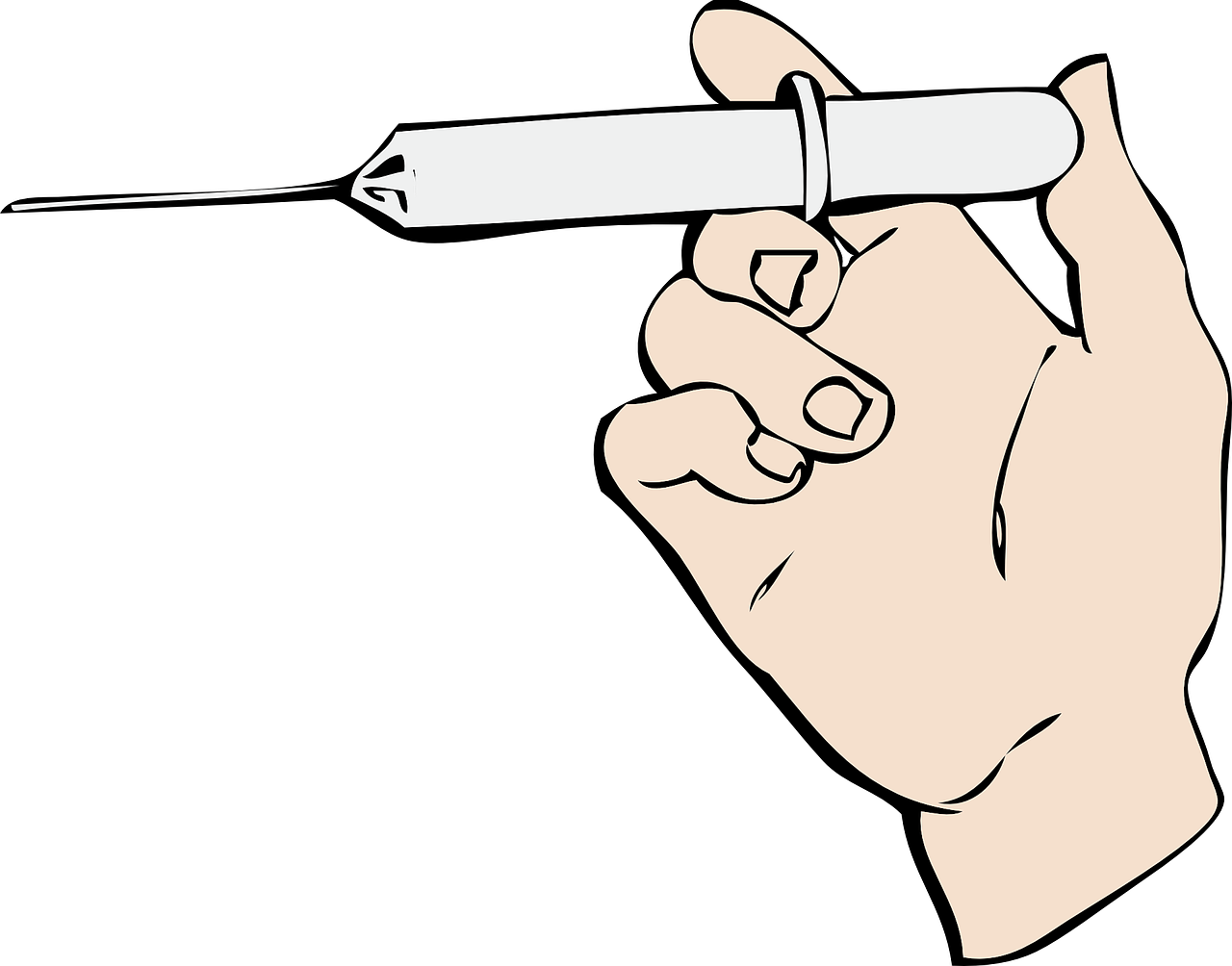}
\quad
\quad
\quad
\includegraphics[width=4cm]{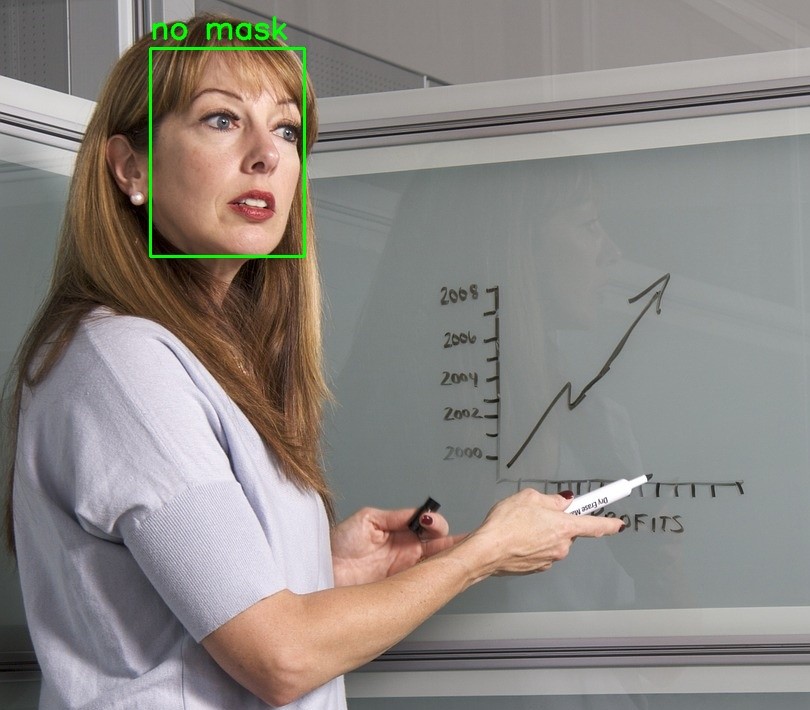}
\quad
\includegraphics[width=4.cm]{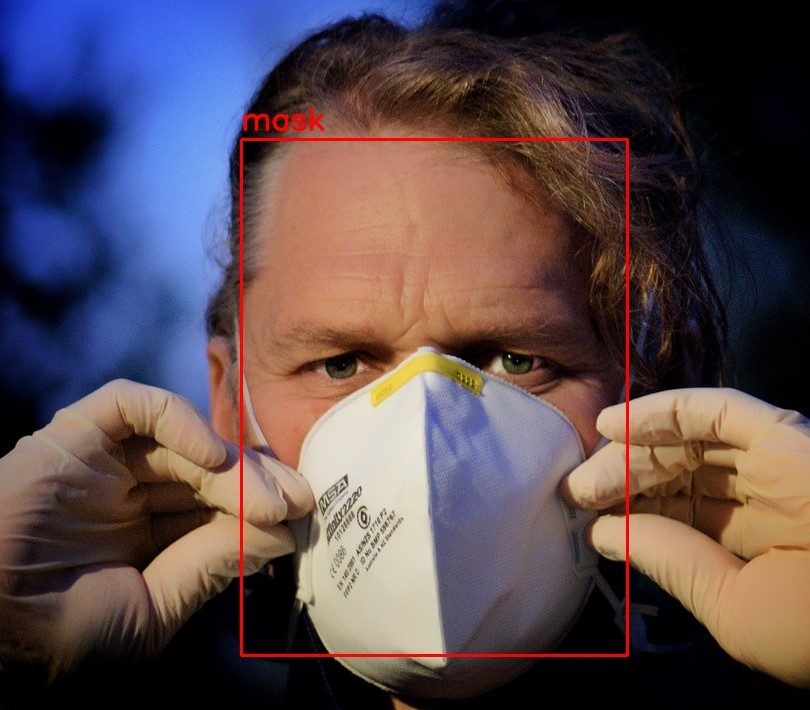}
\includegraphics[width=13cm]{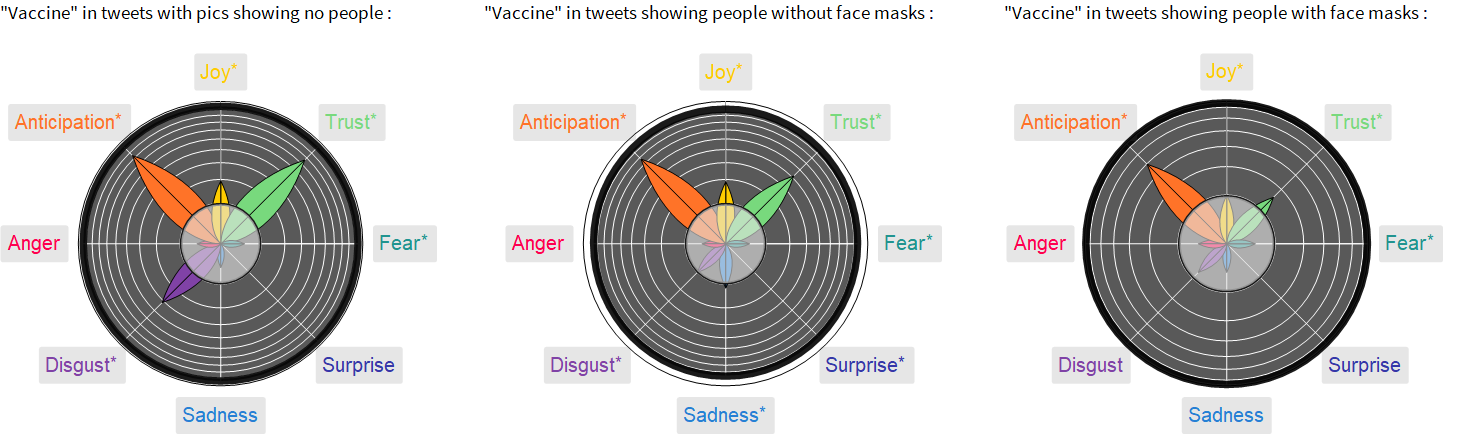}
\includegraphics[width=4cm]{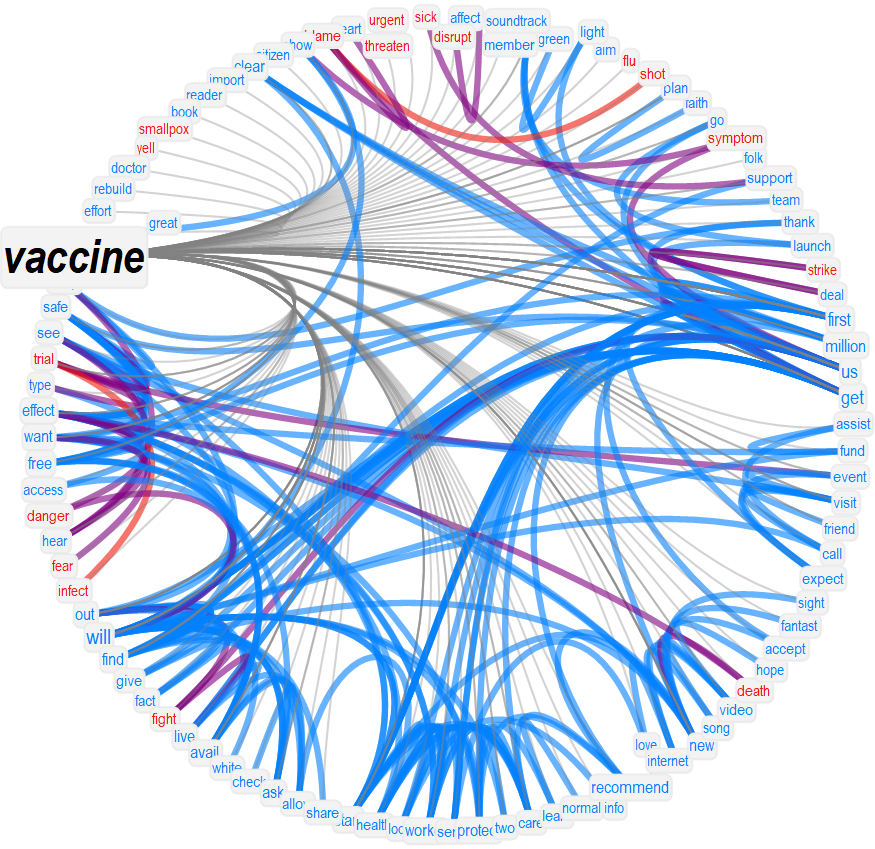}
\includegraphics[width=4cm]{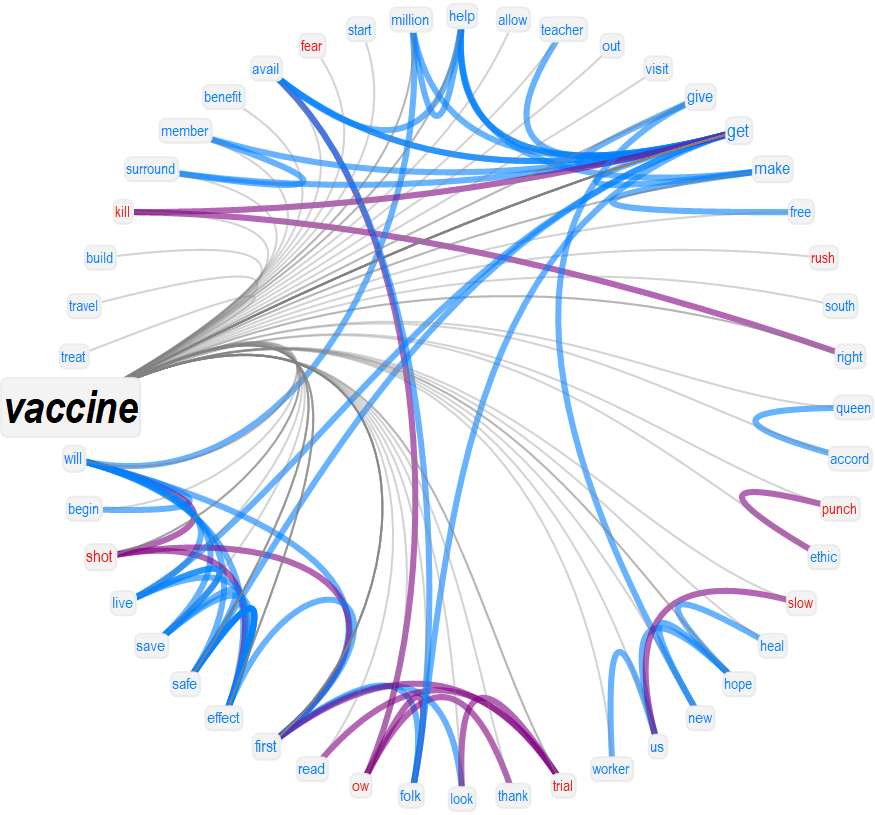}
\includegraphics[width=4cm]{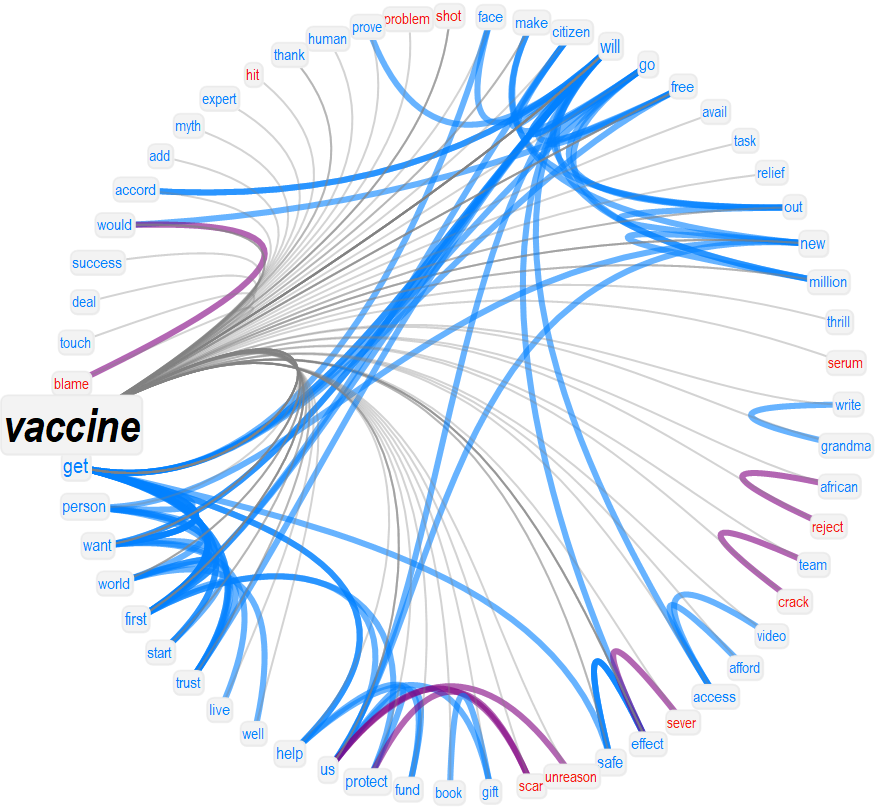}
\caption{Emotional flowers and valenced semantic frames for "vaccine" in those tweets including pictures with: (i) no people (left), (ii) people wearing no face masks and (iii) people wearing face masks. On the top part of the panel there are example pictures that were taken from Pixabay for demonstrating how the implemented Python library works. Bottom: Semantic frames reporting only negative and neutral words associated to "vaccine".}
\label{fig:facemasks}
\end{figure}

The emotional profiles of "vaccine" contained in these three categories of multimedia are reported in Figure \ref{fig:facemasks}. Different emotions are found to populate the semantic frame of "vaccine" across these three categories. The language of popular tweets including pictures with no people is strongly polarised between trust/joy and disgust, emotions that are absent in popular tweets portraying people. 

Tweets showing people wearing face masks corresponded to an emotional profile for "vaccine" different from the one of tweets showing people without masks. Pictures showing the full face of a person were accompanied with more trustful and slightly more joyous language - when associated to the idea of the vaccine - in comparison to the messages accompanying tweets with people wearing face masks. No difference was found in terms of anticipation, which permeates all the considered semantic frames of "vaccines" in Figure \ref{fig:facemasks}. The combination of anticipation, trust and joy in popular tweets with people wearing no face masks is a marker for hopeful emotional states, projected with positive affect into the future. This pattern is confirmed by the association between "vaccine" and "hope" in the respective semantic frame (see Fig. \ref{fig:facemasks} bottom). This hopeful framing vanished in the language of messages reporting people wearing face masks.

\subsection{Aftermath of the temporary suspension of AstraZeneca's vaccine: Loss of trust in the Italian twittersphere}

On March 15th 2021, several European countries, including Italy, temporarily suspended the use of the COVID-19 vaccine developed by AstraZeneca, following sparse reports of serious side effects. 

\begin{figure}[]
\centering
\includegraphics[width=9.cm]{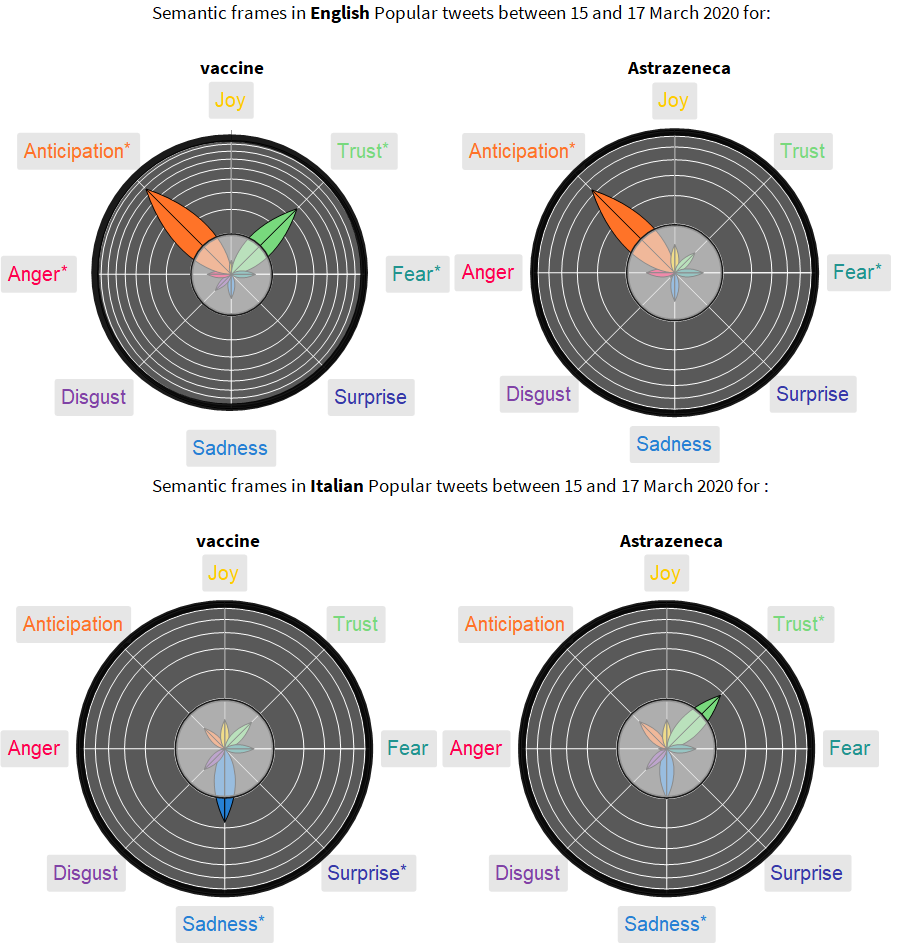}
\caption{Emotional flowers for "vaccine" and "astrazeneca" in popular tweets gathered after the suspension of the AstraZeneca vaccine in several EU countries in mid March 2021. These results should be compared with the emotional profiles reported in Figure \ref{fig:vacci} and relative to months before the suspension.}
\label{fig:aftermaths}
\end{figure}

Figure \ref{fig:aftermaths} reports the emotional profiles of "vaccine" and "astrazeneca" in popular tweets gathered in the next few days after the suspension. In comparison with the popular perceptions observed in December 2020 and in January 2021, as summarised in Figure \ref{fig:vacci}, the temporary suspension of the AstraZeneca vaccine had drastic effects over social discourse in the Italian twittersphere but not in the English one.

English popular tweets still framed the idea of the vaccine along strong signals of anticipation into the future and trust. Trust was found in the semantic frame of "vaccine" but not with regards to the syntactic/semantic associates of "astrazeneca", indicating a potential shift in trust between the general concept of a COVID-19 vaccine and the concrete one by AstraZeneca's as described in popular tweets.

A more drastic shift in the emotions of vaccines was found in the Italian twittersphere. By comparing Figures 1 (bottom) and \ref{fig:aftermaths} (bottom) one notices how trust, joy and anticipation expressed by Italian users when mentioning "vaccine" in December and January vanished completely in the aftermath of the AstraZeneca temporary suspension in mid March 2021. Positive emotions disappeared from the semantic frame of "vaccine" and were replaced with a weak signal of sadness, indicating concern as expressed by Italian users in popular messages. 

Notice how sadness permeated the semantic frame of "vaccine" but not the language surrounding "astrazeneca" in Italian, which was rather associated with concepts mildly eliciting trust. A human coding of the popular tweets revealed that the detected trustful language was mainly relative to epidemiology experts trying to convey the relevance of the vaccination programme and the unfairness of incorrect statistical biases towards vaccination and its risks as portrayed by news media.

\section*{Discussion}

This work investigated social media language around COVID-19 vaccines. Focus was given to popular messages on the Twittersphere, i.e. content identified by the platform as being highly re-tweeted and liked by online users. 

Social media provide crucial data for understanding how large audiences perceive and react to events \cite{ferrara2015quantifying,stella2021cognitive}. Past approaches used social media traffic for inferring electoral outcomes in massive voting events \cite{bessi2016social,bovet2018validation} and more recent approaches adopted social media language to understand how massive populations coped with the COVID-19 pandemic \cite{dyer2020public,stella2020lockdown,aiello2020epidemic,montefinese2021covid}. The overarching theme of these approaches, including the current one, is that language is a driver of emotional content and conceptual knowledge that is transferred from people's minds into digital posts \cite{stella2021cognitive}. Accessing and modelling knowledge in social discourse becomes thus a proxy for reconstructing how massive audiences framed - semantically and emotionally - events in their cognitive representations of the world \cite{fillmore2006frame}. In particular, understanding how popular posts frame ideas and emotions is crucial because popularity can lead to one single tweet being read and endorsed by up to 500k users. While not every user might be human \cite{stella2018bots,gonzalez2021}, these numbers indicate how crucial popular content can be in influencing people's perceptions of real-world and online events, as confirmed also by recent works \cite{bessi2016social,bovet2018validation}.

Popular tweets were found to portray mainly logistic aspects of vaccine distribution. Frequent and semantically rich \cite{vitevitch2019can} concepts of social discourse were relative to the necessity for people to receive as soon as possible their first dose of vaccine despite the issues of administering massive amounts of vaccine. 

English discourse in popular tweets was found to be strongly emotionally polarised. Emotional profiles featured at the same time strong signals of trust/joy and sadness/anger. Negative emotions were found to be channelled in language denouncing the issues of vaccine distribution rather than in the pandemic by itself, differently from what was found in the early stages of the pandemic \cite{dyer2020public,aiello2020epidemic,montefinese2021covid,stella2020lockdown}. Trust and joy were rather relative to jargon celebrating science and its success in delivering a tool for fending off the pandemics. As a future research direction it would be interesting to identify whether these opposing emotions were expressed by the same set of users over time or were symptoms for the creation of different topics coexisting in the online information flow (like detected in \cite{ferrara2015quantifying} and more recently in the socio-semantic analysis by \cite{radicioni2021networked}).

Almost no emotional polarisation was found in the semantic frame of "vaccine" in Italian, where an excess of positive emotions like joy and trust were found, in addition to anticipation. According to Ekman's atlas of emotions \cite{ekman1994nature}, these basic emotions in language can give rise to nuances like positive expectations for the future, i.e. hope. 

Regretfully, the hopefulness expressed by Italians in popular tweets vanished completely after the case of the temporary suspension of the AstraZeneca vaccine in Italy and in several other EU countries. Our retrieved emotional profiles identified a drastic shift in the emotional portrayal of vaccine in online popular tweets. We registered a transition from a trustful and hopeful perception to a subsequent connotation of "vaccine" permeated by sadness, with no anticipation, joy or trust surrounding it. This is a concerning finding because of increasing psychological evidence that a lack of trust towards the health system and the institutions regulating is a marker for a reluctance to adopt health-related social norms like vaccination \cite{murphy2021vaccines,kalimeri2019human}, with concrete negative repercussions for global health. 

A key innovation of our approach was combining cognitive networks of linguistic associations \cite{siew2019cognitive,stella2020text} together with pictures and AI-based methods of image analysis \cite{reece2017instagram,deng2019retinaface}. In addition to evidence of emotional coherence between the text embedded in pictures and the one coming from tweets, our work identified some differences in the language accompanied by specific categories of pictures. English popular tweets showing no pics of people were found to frame the idea of "vaccine" along with contrasting emotions of trust and disgust. A semantic network analysis for these emotional profiles identified their source in discussion focused on potential side effects and, more importantly, in associations reporting vaccines as ways to mitigate the impact of COVID-19. 

Language accompanying tweets with people wearing a face mask exhibited almost no signal of trust or joy in the semantic frame of "vaccine", differently from the portrayal of vaccines produced by messages including pictures with people wearing no face mask. Although co-occurrence does not imply causation, it must be noted that the adoption of face masks in public places has been met with mixed results (for a review see \cite{perra2021non}) by most Western countries. Masks hinder expressiveness and can create discomfort if worn for long times but there are also additional psychological elements. In fact, recent psycholinguistic studies showed how face masks became strongly associated with negative concepts like sickness and disease in the cognitive perception of the COVID-19 pandemics \cite{mazzuca2021conceptual}. In this way, by merging pictures and social media discourse, our results indicate that the overall perception of people wearing face masks is biased, i.e. poorer in terms of trust and joy when compared to common portrayals of people wearing no face protection. Future research should explore the emotional profiling of face masks and other aspects of COVID-19 vaccines on larger scales.

We also identified a behavioural tendency for English users to re-share more emotionally extreme content, i.e. featuring stronger signals of disgust, sadness and joy, but also to favour less content enticing more sadness. These patterns were absent in the Italian twittersphere. The interplay between emotions and content re-sharing extends previous results related to sentiment polarity \cite{ferrara2015quantifying} which highlighted a positive bias in content sharing and liking, i.e. positive sentiment promoting content sharing and endorsement. Our findings indicate that also negative, inhibiting emotions like disgust can amplify content sharing while sadness inhibits endorsement of online posts.

Last but not least, our approach identified concerning associations between vaccines and conspiratorial jargon. Conceptual associations between hoaxes and vaccines have been traced in many other studies, cf. \cite{kalimeri2019human}. However, in this case these associations were found in popular messages and not in borderline peripheral content, like the one produced by malignant social bots \cite{stella2018bots,gonzalez2021}. A negative framing of vaccines in terms of hoaxes driven by popular messages to large populations could self-evidently have negative consequences for the vaccination campaign. A recent study from cognitive neuroscience found that conspiracy-like misinformation can decrease pro-vaccination attitudes by exploiting the emotion of anger \cite{featherstone2020feeling} (an emotion detected here) rather than through fear (which was not detected here). Anger can activate and amplify reactions like feeling frustrated or fooled by the establishment, which then lead to behavioural changes \cite{ekman1994nature}. In this way both the conspiratorial semantic associations and emotional signal of anger in the semantic frame of "vaccine" in popular English tweets should serve as early warning signals of misinformation hampering the vaccination campaign. 

Notice that our analysis is subject to some limitations. Textual forma mentis networks can adapt their structure but not their valence/emotional labels to the text being analysed. This is because affect data comes from predetermined large populations and it is representative of the way common language portrays concepts \cite{warriner2013norms,mohammad2016sentiment,stella2020text}. Because of contextual shifts, it might be that the affective connotation of specific concepts in a given discourse could be different \cite{stella2019innovation}. For instance, "vaccine" by itself was rated as mostly neutral in the dataset by Warriner and colleagues \cite{warriner2013norms} but it was associated with mostly positive jargon in social discourse (see Figure \ref{fig:vacci}, bottom). This limitation underlines the importance of considering words as connected with each other and not in isolation. This is because TFMNs enable the reconstruction of contextual shifts in affect by considering how words were associated with each other in language.  Another limitation of the current approach is that the multi-language support stems mainly from automatic translation. Hopefully in the future more cognitive datasets will be built by considering native speakers' data. 

As a future research direction it would be interesting also to merge the current semantic-emotional analysis together with content credibility dynamics, as recently quantified in social discourse about the vaccine by Pierri and colleagues \cite{pierri2021vaccinitaly}.

\section*{Conclusions}

Our work provides a methodological framework for reconstructing trending perceptions in social media via language, network and picture analyses. Warning signals of conspiratorial content and a dramatic loss of trust towards the vaccination campaign were unearthed by our investigation. Our results stress the possibilities opened by innovative and quantitative analyses of semantic frames and emotional profiles for understanding how large populations of individuals perceive and discuss events as extreme as the global pandemic and ways out of it like vaccines.

\section*{Acknowledgements}

The authors acknowledge Riccardo Di Clemente for insightful discussion.

\bibliographystyle{unsrt}
\bibliography{sample}

\newpage

\section*{SUPPLEMENTARY INFORMATION}

This section provides additional information and quantitative evidence supporting the main text.

\begin{figure}[ht]
\centering
\includegraphics[width = 0.4\textwidth]{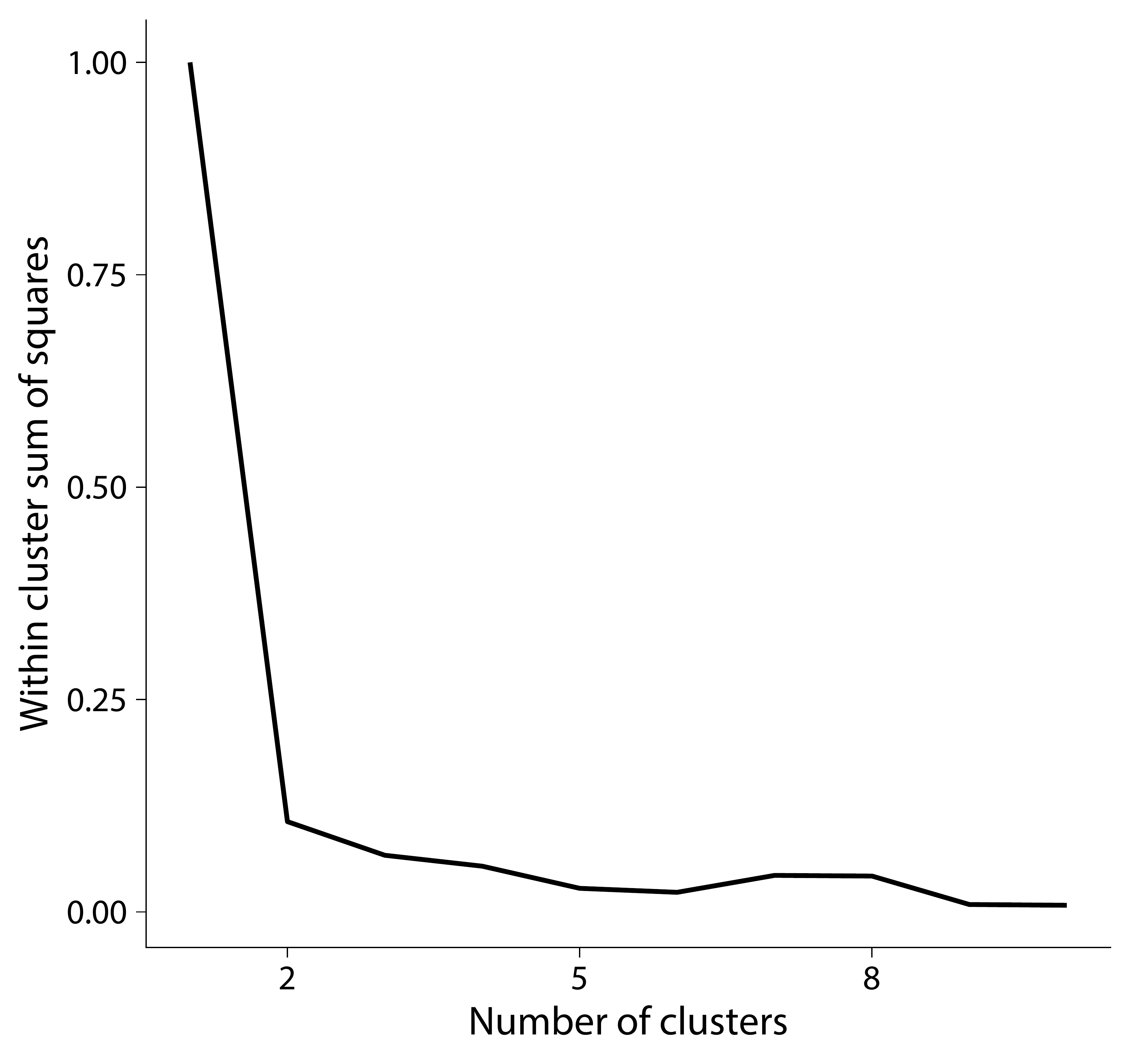}
\quad
\includegraphics[width = 0.4 \textwidth]{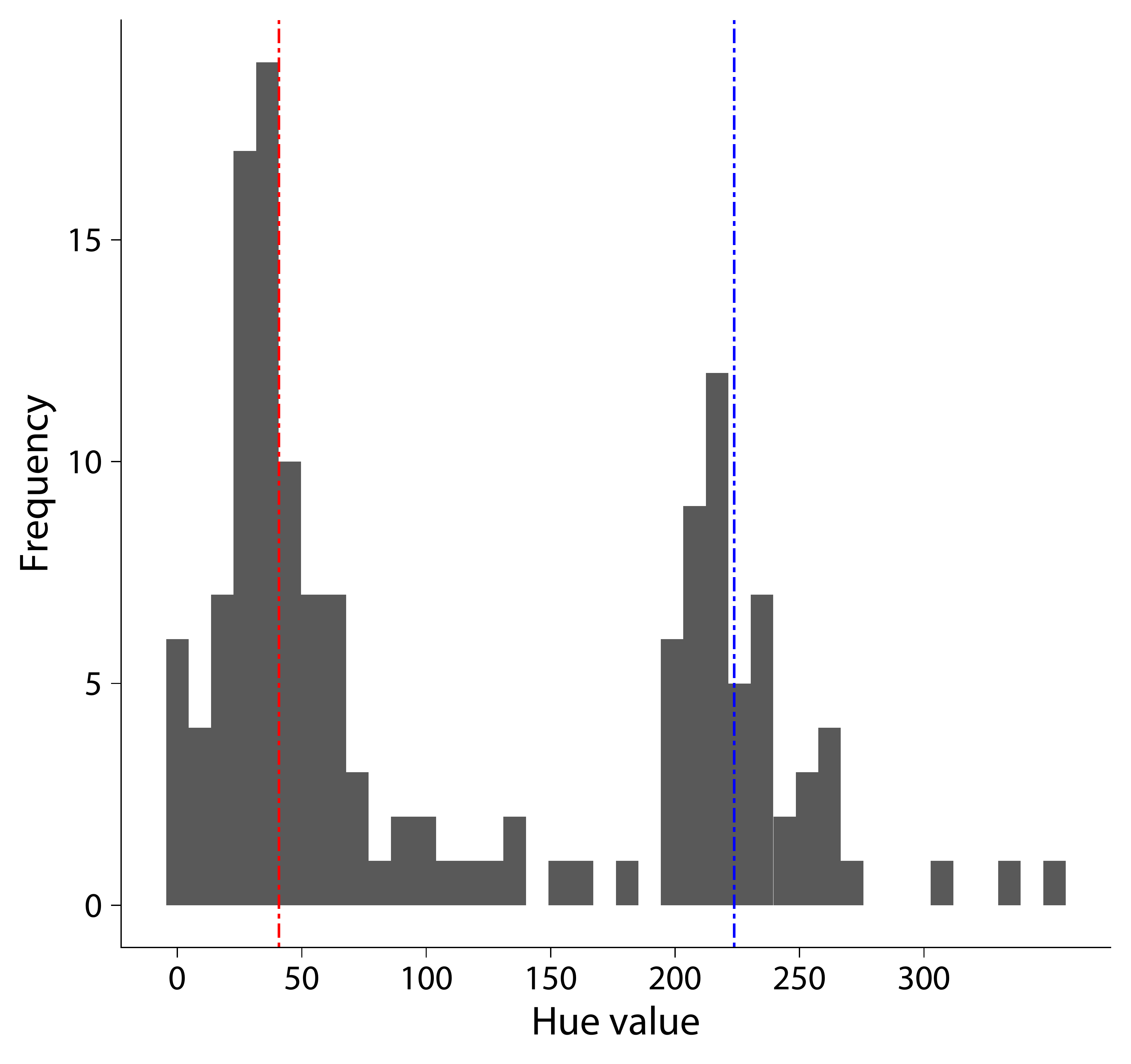}
\caption*{\textbf{Supplementary Figure 1} - Left: elbow plot showing how the within cluster sum of squares varies as the number of clusters increase, when clustering the images based on their dominant hue values. As the plot indicates, two clusters seem to be the optimal choice. Right - histogram of the dominant hue values detected from the images (note: only those not containing any text were analysed in this scenario). As visual inspection suggests, we observe two clusters centred on the red and blue tones of hue.}
\label{fig:supp_clustering}
\end{figure}

\begin{figure}[ht]
\centering
\includegraphics[width=10cm]{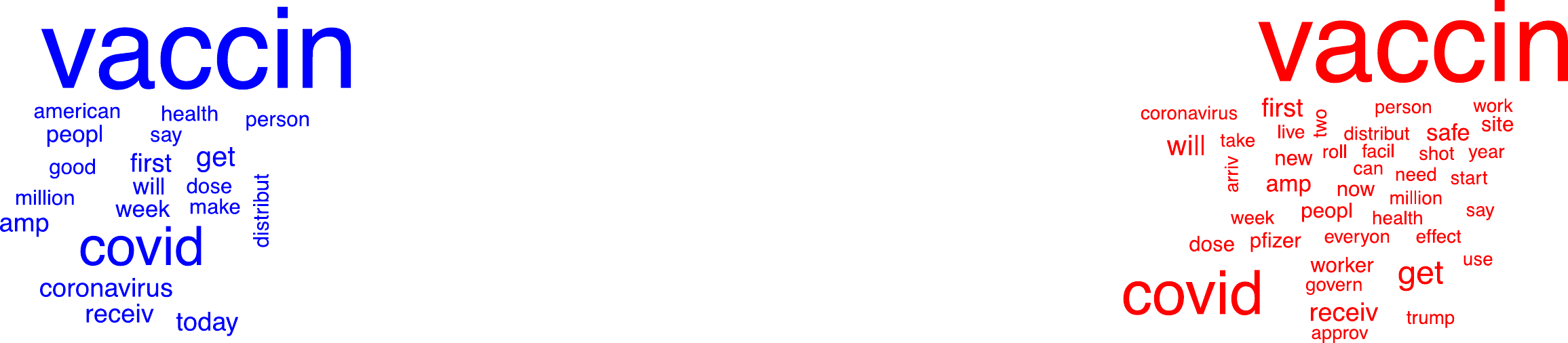}
\includegraphics[width=10.5cm]{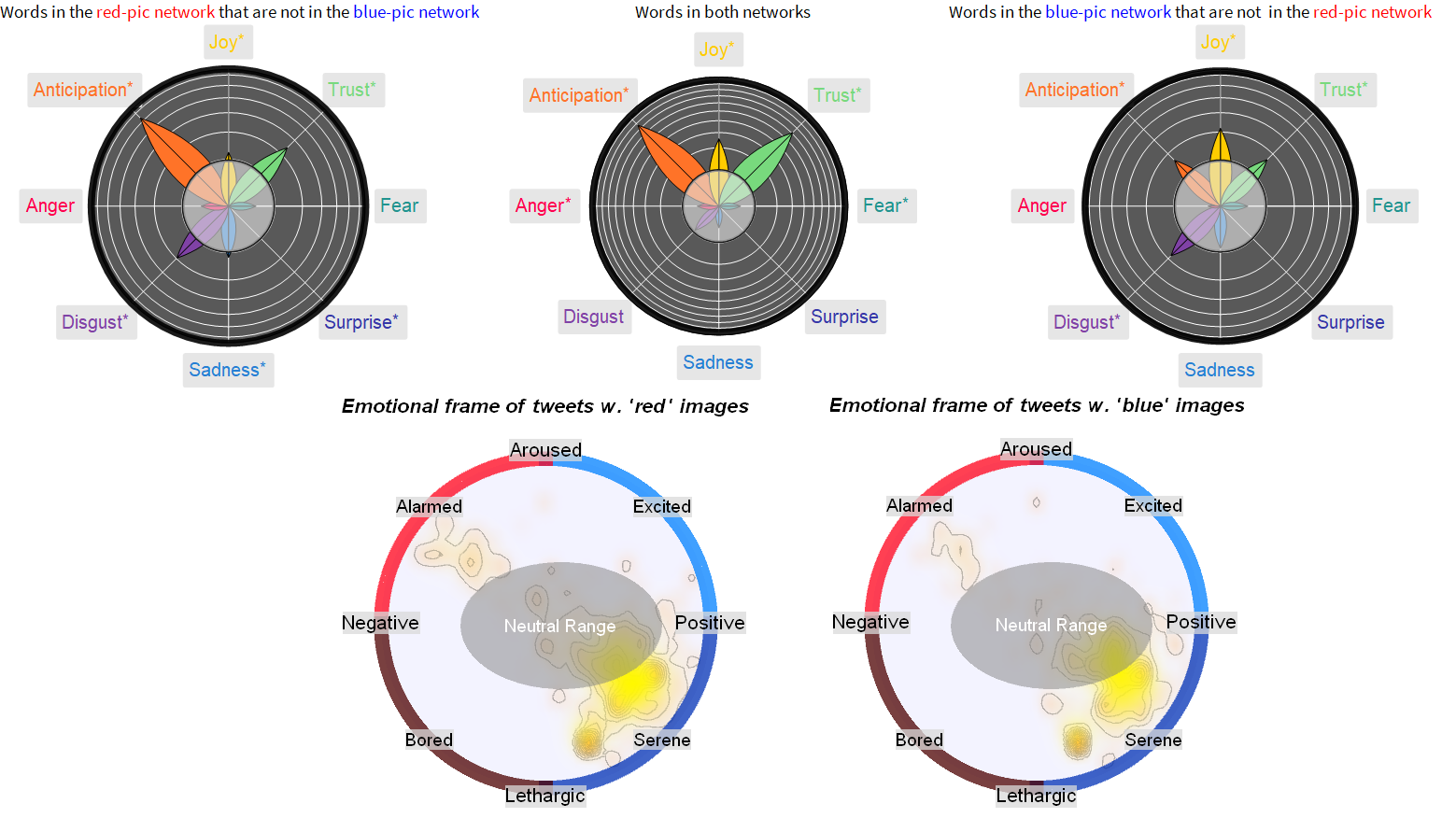}
\caption*{\textbf{Supplementary Figure 2} - Top: Word cloud of the most frequent words in tweets with pictures with blue or red as predominant. Bottom: Emotional flowers and circumplex model for the emotions of the language used in tweets with pics of different predominant colours.}
\label{fig:redandblue}
\end{figure}

\end{document}